# Flow cytometry to assess the counts and physiological state of *Cronobacter sakazakii* cells after heat exposure

De-La-Cal-Sabater, P.[1,2], Caro, I.[1], Mateo, J.[3], Castro, M.J.[2], Cao, M.J.[2], Quinto, E.J.[1]*

[1] Department of Nutrition and Food Science, Faculty of Medicine, University of Valladolid, 47005, Valladolid, Spain.

[2] Department of Nursery, Faculty of Nursery, University of Valladolid, 47005, Valladolid, Spain.

[3] Department of Food Hygiene and Food Technology, Faculty of Veterinary Medicine, Campus de Vegazana s/n, 24071, León, Spain.

De-La-Cal-Sabater, P. (calsabater89@gmail.com); Caro, I. (icarc@unileon.es); Mateo, J. (jmato@unileon.es); Castro, M.J. (mjcasalija@gmail.com); Cao, M.J. (mjcao@enf.uva.es); Quinto, E.J. (equinto@ped.uva.es)

* Author for correspondence: E.J. Quinto. Department of Food Science and Nutrition, Faculty of Medicine, University of Valladolid, 47005, Valladolid, Spain. E-mail: equinto@ped.uva.es.

ABSTRACT

*Cronobacter sakazakii* is an opportunistic pathogen associated with outbreaks of neonatal necrotizing enterocolitis, septicemia, and meningitis. Reconstituted powdered infant formulae (PIF) is the most common vehicle of infection. Plate count methods do not provide direct information on the physiological status of cells. Flow cytometry (FC) has been used to gain insights into the physiological states of *C. sakazakii* after heat treatments, and to compare FC results with plate counts. The percentage of compromised cells increased as the percentage of live cells increased after the 100 ºC treatment. However, the number of compromised cells after 60 or 65 ºC treatments decreased as the percentage of live cells increased, showing that both mild temperatures would not be completely effective eliminating all bacteria but compromising their membranes, and showing that mild heat treatments are not enough to guarantee the safety of PIF. FC was capable to detect *C. sakazakii* compromised cells that cannot be detected with classical plate count methods, thus it could be used to decreasing the risk of pathogenic viable but non-culturable cells to be in the ingested food. Linear regression analysis showed good correlations between plate count results vs FC results.

1. Introduction

During the last decade, scientific interest has turned to *Cronobacter sakazakii*, formerly *Enterobacter sakazakii*, as a human pathogen. *C. sakazakii* is an opportunistic pathogen associated with outbreaks of neonatal necrotizing enterocolitis, septicemia, and meningitis (Beuchat et al., 2009; Farmer, Asbury, Hickman, & Brenner, 1980; Iversen et al., 2008). The first virulence factors identified in *C. sakazakii* were enterotoxins (Pagotto et al., 2003). Raghav & Aggarwal (2007) identified a 66kDa toxin which was most active at pH 6, with the ability to be stable at 90 ºC for 30 min, and potent cell toxicity ($LD_{50}$ = 56 pg). This organism has a high case fatality rate in vulnerable infants and, in surviving patients, severe neurological sequelae have occurred including hydrocephalus, quadriplegia and developmental delay (Iversen et al., 2008; Lai, 2001).

The International Commission for Microbiological Specification for Foods (2002) ranked *Cronobacter* spp. as a "severe hazard for restricted populations causing life-threatening or substantial chronic sequelae of long duration." The FAO/WHO (2004) noted that although *C. sakazakii* has caused invasive infection in all age groups, infants are the group at particular risk. Reconstituted powdered infant formulae (PIF) and powdered milk have been the most common vehicles implicated in neonatal *C. sakazakii* infections (Beuchat et al., 2009; Biering et al., 1989; Centers for Disease Control and Prevention, 2002; Simmons et al., 1989; van Acker et al., 2001).

PIF could be easily contaminated because it is a non-sterilized product (Beuchat et al., 2009). In hospitals, environmental contamination and temperature abuse of the reconstituted formula have been contributory factors (Bar-Oz et al., 2007; Biering et al., 1989). The WHO (2007) recommends cooling the temperature of boiled water not below 70 ºC for safe preparation of PIF. Heat stress can damage the bacterial cell wall and cause the breakage of genomic DNA along with misfolding of cytoplasmic proteins (Arku, Fanning, & Jordan, 2011a). However, bacteria can survive to sublethal heat stress and can also be adaptive by inducing response to high environmental temperatures resulting in higher tolerance to subsequent lethal stress (Jordan et al., 1999), which has been demonstrated for *Cronobacter* spp. (Arku, Fanning, & Jordan, 2011b).

The plate count-based method to assess bacterial numbers surviving after the imposition of stress does not provide direct information on the physiological status of cells (Arku et al., 2011a; Herrero, Quirós, García, & Díaz, 2006). This approach only detects cells able to form colonies but cannot detect metabolically active cells that do not form them (Bunthof and Abee, 2002; Herrero et al., 2006), such as compromised – damaged or injured– or stressed bacteria, which could be viable but nonculturable (Arku et al., 2011; Baatout et al., 2005; Kim et al., 2009). In contrast, flow cytometry (FC) provides information on the dynamics and physiological heterogeneity of bacterial populations (Amor et al., 2002; Pianetti et al., 2005). This technique employs fluorescence markers and uses direct optical devices to quantify the properties of single cells (Papadimitriou et al., 2006), providing a fundamental advantage over conventional methods (Nebe-von-Caron et al., 2000; Shapiro, 2000).

The aim of the present study is to apply the FC technique to *Cronobacter sakazakii* cultures after mild and strong heat treatments to gain insights into bacteria physiological states, and to evaluate the performance of the FC for the quantification of *Cronobacter sakazakii* as compared with plate counts.

## 2. Material and methods

*2.1. Culture preparation*

*Cronobacter sakazakii* ATCC 29544 was used (Choi, Kim, Lee, & Rhee, 2013; Joshi, Howell, & D'Souza, 2014; Lehner et al., 2005; Parra-Flores, Juneja, de Fernando, & Aguirre, 2016). In triplicate (in three different days) the strain was cultured twice in sterile Trypticase Soy Broth (TSB; Difco, BD Diagnostics, Spark, MD, USA) at 37 ºC for 24 h to reach the stationary phase, with a concentration of ca. $10^9$ CFU/ml. The populations were counted after spreading 10 µl aliquots from serial dilutions onto TSA plates following the drop method described by Chen, Nace, & Irwin (2003) and incubated at 37 ºC for 24 h. A population of $1.1 \times 10^9$ ($\pm 0.1 \times 10^9$) CFU/ml was counted.

*2.2. Dead-cells suspensions (DCS)*

Dead-cells single-color suspensions were prepared as controls for setting up the FC procedure following the manufacturer instructions of the staining kit used later (see below). Portions of 1 ml from the microorganism's cultures were dispensed into microcentrifuge tubes, centrifuged at 10,000 *g* for 3 min to pellet the cells and washed with 0.85% NaCl. The supernatants were removed and discarded. Centrifugation and washing procedures were repeated twice. The pellets were re-suspended with 1 ml of 70% isopropyl alcohol and incubated at room temperature for 60 min mixing every 15 min, giving DCS ready for the staining.

*2.3. Non-heat-treated (nTCS) and heat-treated (TCS) cell suspensions*

One ml aliquots from the microorganism's cultures were dispensed into microcentrifuge tubes and centrifuged at 10,000 *g* for 3 min. The pelleted cells were washed with 0.85% NaCl and the supernatants were removed and discarded. Centrifugation and washing procedures were repeated twice. The pellet was re-suspended with 1 ml of 0.85% NaCl giving a non-heat-treated cell suspension (nTCS), and an aliquot was used for viable cell count as mentioned above (Chen, Nace, & Irwin, 2003).

For the heat treatment viability assays, 10 ml from broth cultures were placed in 50 ml screw-cap glass tubes and heated at 60, 65, or 100 ºC for 5 min and immediately introduced in a water bath at 4 ºC for 5 min. As *C. sakazakii* mean thermal inactivation $D_{60 ºC}$ and $D_{65 ºC}$ values are about 4.3 and 0.6 min, respectively (z-value of 5.6 ºC; Edelson-Mammel and Buchanan, 2004), partial thermal inactivation and sublethally damaged cells resembling under-pasteurization heating conditions were expected after both 60 or 65 ºC heat treatments. Irreversibly destroyed cells were expected after the 100 ºC treatment resembling high pasteurization conditions. After the heat treatments, 1 ml of each of the heat-treated cultures were dispensed into microcentrifuge tubes, centrifuged and washed twice with 0.85% NaCl as previously described obtaining heat-treated cell suspensions (TCS); the resulting viable cell concentrations were determined as described for the nTCS.

Finally, cell suspensions were mixed into FC analysis tubes at different nTCS:TCS ratios (vol:vol): 100, 75, 50, 25, and 0% of nTCS. Control tubes with only 0.85% NaCl were also prepared.

*2.3. FC analysis*

Cell suspensions were stained using the Live/Dead BacLight Bacterial Viability and Counting Kit (Molecular Probes, Invitrogen, CO, USA) according to the manufacturer instructions. For accurate counting, the total volume in the FC tubes was 1000 µl. Briefly, 10 µl of the bacterial cell suspension, 987 µl of 0.85% NaCl, 1.5 µl of 3.34 mM SYTO9 green fluorescent nucleic acid stain and 1.5 µl of 30 mM propidium iodide (PI) red fluorescent nucleic acid stain were dispensed into each FC tube. The tubes were incubated at room temperature for 15 min protected from light.

FC analyses were conducted using a Cytomics FC 500 (Beckman Coulter, Brea, CA, USA) with an excitation wavelength of 488 nm from a blue argon laser. Each cell (called event) was characterized by two fluorescent parameters that measured green fluorescence emission (FL1 channel) and red fluorescence emission (FL3 channel). Green fluorescing SYTO9 is able to enter all cells and is used for assessing total cell counts, whereas red fluorescing PI enters only cells with damaged cytoplasmic membranes. Optical bandpass filters were set up to measure the green fluorescence of SYTO9 at 525/40 nm (FL1) and the red fluorescence of PI at 620/30 nm (FL3). The measurements were conducted after 300 seconds or 500,000 events at a low flow rate (10 µl/min).

Correlations between FC method and plate counts were calculated (Gunasekera et al., 2000). Different cell concentrations of *C. sakazakii* between $10^4$ and $10^9$ CFU/ml were analyzed using FC and plate count method.

## 3. Results

*3.1. DCS*

The FC plot obtained from *C. sakazakii* dead-cells suspensions is shown in Figure 1. The events are distributed according to fluorescence intensity at the two wavelengths studied. A gate was positively identified for dead cells since all of them were restricted to a very well-defined area, allowing its use as a control for future differentiation between dead and live cells in different areas.

*3.2. nTCS and TCS*

Figure 2 shows the distribution of the events observed in the cytometer according to the staining of nTCS and different nTCS:TCS ratios. Figure 2A shows three different 100% nTCS plots before been heat treated. An area for the live cells was identified. Figure 2B shows the nTCS:TCS mixtures: 0, 25, 50, or 75% of nTCS cultures mixed with TCS previously heat-treated at 60, 65, or 100 ºC. After the 100 ºC heat treatment, dead cells were restricted to a very well-defined area (Figure 2.B3; 0% nTCS or 100% TCS), as previously observed from DCS controls shown in Figure 1. The identification of live cells and dead cells areas allowed the identification of the third area for cells with compromised membranes –stained with both colorants. Compromised cells spread themselves into the intact-membrane (live cells) and damaged-membrane (dead cells) areas as observed with the different nTCS:TCS mixtures. After the 100 ºC heat treatment, the number of compromised cells increased as the percentage of live cells (nTCS) increased. In contrast, after the heat treatments at 60 or 65 ºC, the number of compromised cells decreased as the percentage of nTCS increased (Figures 2.B1 and 2.B2), showing that mild heat treatments at 60 or 65 ºC do not kill completely all bacteria but part of the population, resulting in a high percentage of compromised cells.

*3.3. FC events vs plate counts*

Linear regression analysis between FC events of live cells and plate counts was carried out obtaining a strong correlation ($R^2 = 0.932$).

## 4. Discussion

*C. sakazakii* is a pathogenic bacterium able to produce capsular material and biofilms (Iversen, Lane, & Forsythe, 2004; Lehner et al., 2005). These characteristics allow the microorganism to protect itself from hostile environments. Three different heat treatments were used to assess the viability of *C. sakazakii* cells, and compromised cells were found. The percentage of compromised cells increased as the percentage of live cells increased after the heat treatment at 100 ºC, and the number of live cells will depend on the environmental conditions affecting their physiological state. Indeed, each cell has its own characteristics, behavior, and response to the environment. Following previous statements by Fredrickson, Ramkrishna, & Tsuchiya (1967) and Baranyi & Roberts (1994) the kinetics of a bacterial population can be characterized by the extracellular environment and the intracellular conditions. This is an interesting fact far from the aim of this work; the study of how and when cells enter a compromised physiological state will need further investigation. The results were different after heat treatments at 60 or 65 ºC, i.e., the number of compromised cells decreased as the percentage of live cells increased; both temperatures were not completely effective at killing all bacteria but compromising their membranes and, subsequently, their viable physiological state.

SYTO9 is a hydrophobic cell-permeant nucleic acid stain that shows a large fluorescence enhancement upon binding nucleic acids after penetrating the cell intact membrane (Bunthof and Abee, 2002; Stocks, 2004). PI is a membrane impermeant hydrophilic stain with a high molecular weight that binds to nucleic acids only when the cell membrane has pores (Manini and Danovaro, 2006; Müller and Nebe-von-Caron, 2010). When the PI is mixed with the SYTO9, cells with intact membrane fluoresce green while cells with damaged membrane fluoresce red –they lose their green fluorescence because the PI competes for the same target areas than the SYTO9, reason behind its greater affinity for nucleic acids than SYTO9 forcing its displacement from its binding (Berney et al., 2007; Hewitt & Nebe-Von-Caron, 2004; Manini & Danovaro, 2006; Stocks, 2004). According to Berney, Hammes, Bosshard, Weilenmann, & Egli (2007), microscopy allows to distinguish the difference between green or red fluorescent cells, but FC allows to observe a curve-shaped pattern of fluorescence with different amounts of both stains. A third green-red fluorescent group of cells can be detected (Pianetti et al., 2005; Subires et al., 2014), which represents an intermediate state in the permeabilization of the cell membrane allowing the PI to penetrates into the cell but not in enough quantities to displace efficiently the SYTO9 from its binding to the nucleic acids (Barbesti et al., 2000; Ben-Amor et al., 2005; Pianetti et al., 2005; Saegeman et al., 2007; Subires et al., 2014).

To study the physiological states of the bacterial cells, SYTO9 and PI stains enable differentiation between bacteria with intact and damaged cytoplasmic membranes, and it has been used to differentiate between active and dead cells (Berney et al., 2007; Gasol et al., 1999; Sachidanandham et al., 2005). Several authors have shown that the staining of bacterial cells with SYTO9 and PI does not always produce distinct live and dead populations (Sachidanandham et al., 2005); indeed, intermediate states reflecting physiological heterogeneity are also observed (Barbesti et al., 2000; Berney et al., 2007, 2006; Gregori et al., 2001; Hoefel et al., 2003; Joux and Lebaron, 2000; Pianetti et al., 2005; Sachidanandham et al., 2005; Virta et al., 1998). As previously stated by Berney et al. (2007), the region of intermediate states shown by the FC plots is referred to as "unknown" in the kit manufacturer's manual also used in the present study, leading to difficulties in the interpretation of results. That fact can be critical when decisions have to be made about the number of viable bacteria in water or food samples (Berney et al., 2007; Pianetti et al., 2005).

To our knowledge, the nature of this intermediate state has not been clarified or detected in *C. sakazakii*. In the present study, we have detected such an intermediate state and linked it to the physiological properties of the bacterial cells. We have observed a distinctive curve-shaped pattern of fluorescence for *C. sakazakii* cells that were exposed to heat treatments. The cloud of live cells showed a displacement of events from the area of the live cells to the dead cells area following a curve-shaped pattern. This displacement would reflect the different physiological states of Gram-negative bacteria as the damage process of the cells proceeds. A first step would be the damage of the outer membrane while the cytoplasmic membrane is still intact –PI cannot penetrate it–, followed by a second step with the complete damage of the cytoplasmic membrane. Berney et al. (2007) reported a similar pattern when Gram-negative bacterial cells such as *Escherichia coli, Salmonella enterica* serovar Typhimurium, and *Shigella flexneri*, were UVA-irradiated or EDTA-treated. That pattern was related to the presence of intermediate cellular states characterized by the degree of damage afflicted specifically on the bacterial outer membrane. These results indicate that the outer membrane of late stationary-phase cells of Gram-negative bacteria is a barrier for SYTO9 (Berney et al., 2007). The permeabilization of the outer membrane can be done with artificial UV (Berney et al., 2007, 2006; Chamberlain and Moss, 1987; Wagner and Snipes, 1982) as well as sunlight (Berney et al., 2006) before the disruption of the cytoplasmic membrane. Our results would show that the disruption of the outer membrane by mild heat treatments may have gone along with some grade of permeabilization of the cytoplasmic membrane; this fact would explain that this phenomenon was not observed in Gram-positive *Enterococcus faecalis* (Berney et al., 2007), which lacks an outer membrane.

*C. sakazakii* may be present in reconstituted PIF and possibly survive the mild heat stress associated with its reconstitution (Parra-Flores et al., 2016) because clinical strains appeared to be more thermotolerant than their environmental counterparts (Yan et al., 2012). Cross-contamination can occur at any point, so those cells present in reconstituted PIF may not be heat damaged (Parra-Flores et al., 2016). Recommendations for the preparation of PIF in the neonatal intensive care units and how to prepare PIF for bottle-feeding at home were proposed by the WHO (2007). To avoid the risk of burns, the appropriate amount of boiled water has to be allowed to cool but not below 70 °C. To achieve this temperature, the water should be left for no more than 30 min after boiling. Despite these recommendations, instructions for reconstitution may suggest using water at temperatures as low as 40 °C (Parra-Flores et al., 2016; Xu et al., 2015) due to undesirable effects on the organoleptic, nutritional, and functional properties of reconstituted PIF (Pina-Pérez et al., 2016). Moreover, the possibility of (i) a re-heating treatment of the previously reconstituted PIF or (ii) a mild heat treatment (40-70 °C) of the water previously to its use for the reconstitution of PIF should not be discarded, especially at home. These situations would represent high-risk scenarios in food safety for children. In the present study, three heat treatments have been used. The 100 °C heat-treated samples did not show bacterial growth on agar plates nor with the cytometer. However, treatments at 60 or 65 °C did not eliminate all bacteria giving counts of about 4-5 log CFU/ml in agar plates and 5.1-5.6 log events/ml with the cytometer. These results show that mild heat treatments are not enough to guarantee the safety of PIF and stress the importance of proper boiling treatment of water previously to its use for reconstituting PIF followed by a cooling period under sterilized conditions until an adequate temperature for consumption is achieved. Parra-Flores et al. (2016) arrived at the same conclusion after studying the variability in cell response of *C. sakazakii* after mild heat treatments at 50 ºC for 5 or 10 min. Arku, Fanning, & Jordan (2011a), Arku, Fanning, & Jordan (2011b), Baatout, De Boever, & Mergeay (2005), and Bunthof & Abee (2002) stated that the membrane integrity reflecting the viability of the cell appears to be dependent on environmental conditions and the physiological status of the cell during the time of analysis; Arku et al. (2011b) showed that *C. sakazakii* cells decrease in survival as the temperature increased from 52 to 67 ºC for 30 min. Choi, Kim, Lee, & Rhee (2013) analyze the mode of membrane disruption with FC founding that the proportion of *C. sakazakii* cells in the PI fluorescent region increased as a function of the treatment time for 5, 10, or 30 min with antimicrobial natural compounds. Marty et al. (2013) studied the molecular dynamics adaptation of microorganisms' proteome to temperature; the dynamic state of a large fraction of the proteome of *Halobacterium salinarum* was strongly disrupted under heat stress from 40 to 60 ºC for 60 min. Similar results were found by Tehei et al. (2004) for *Escherichia coli*.

Parra-Flores et al. (2016) observed considerable variability in the lag phase of *C. sakazakii* cells, with great influence of a previous heat-shock and the growth temperatures. Both findings agree with similar results in previous studies with *Listeria innocua, Enterococcus faecalis, Salmonella enterica*, and *Pseudomonas fluorescens* (Aguirre et al., 2013, 2009; Koutsoumanis, 2008; Lianou and Koutsoumanis, 2011; Metris et al., 2008; Xu et al., 2015).

Compromised cells have special growth requirements due to their physiological state and include two types of cells, i.e., those that could be culturable through proper culture conditions and cells viable but non-culturable. To detect the former it is necessary to adapt the culture conditions such as media, temperature and time (Subires et al., 2014). The most common strategy is to include a recovery step using a non-selective medium previously to the inoculation of the bacteria into the selective media. Some authors propose to use a plate count technique with both non-selective and selective culture media; indeed, the difference between both counts would reflect the count of cells with sublethal damage (Kell et al., 1998; Wesche et al., 2009). Although non-selective methods are used, compromised viable but non-culturable cells cannot be detected by using classical plate count methods (Braux et al., 1997; Gregori et al., 2001; Zhao et al., 2017) and, unfortunately from the public health point of view, they can retain their pathogenic potential. Bacteria culture techniques are time-consuming and do not show the physiological state of the cells (Lemarchand et al., 2001; Ueckert et al., 1997). Recently, Subires et al. (2014) detected the presence of compromised cells of *Escherichia coli* O157:H7 in ready-to-eat pasta salad and its ability to recover along with the refrigerated storage. The present work is not focused on recovering the compromised cells for checking their cultivability with classic plate count methods but detecting them with a fast and reliable technique. FC is a technique that analyses cell by cell quickly and precisely, and it allows to differentiate several types of cells or populations depending on the combination of parameters, such as light and fluorescence, giving us quantitative information about the heterogeneity of the physiological state of bacteria (Davey, 2002).

*3.5. Conclusions*

    *C. sakazakii* may be present in reconstituted PIF and survive the mild heat stress associated with its reconstitution; moreover, cross-contamination with non-heated and compromised cells can occur at any point. Despite the international recommendations for the preparation of PIF, some situations could arise especially at home such as the use of water temperatures as low as 40 ºC or after a mild heat treatment (40-70 ºC)

previously to its use for the reconstitution of PIF, or even a mild re-heating treatment of a previously reconstituted PIF. All these situations increase dramatically the public health risk associated with the presence of *C. sakazakii* in reconstituted PIF. In the present study, three heat treatments have been used. As expected, the 100 ºC heat treatment is the most effective killing the microorganism, allowing the proper assessment of the damaged population with the cytometer; from that assessment onwards, the cytometric technique was able to differentiate the compromised –viable but non-culturable– cells creating a well-defined new cytometric gate vs the live cells populations. Heat treatments at 60 or 65 ºC did not eliminate all bacteria giving high numbers of compromised cells, showing that mild heat treatments are not enough to guarantee the safety of PIF. The importance of proper boiling treatment of water previously to its use for reconstituting PIF must be emphasized. Flow cytometry is able to detect *C. sakazakii* compromised cells that cannot be detected with classical plate count methods. Classic culture techniques are time-consuming and do not show the physiological state of the cells, allowing viable but non-culturable cells maintain their pathogenic characteristics along with the reconstitution of PIF and its consumption by the children.

**Acknowledgments**

This project was financially supported by the I+D+I Program, Consejería de Sanidad, Junta de Castilla y León, Spain (SAN196/VA07/07, SAN673/VA05/08, and SAN126/09).REFERENCES

Aguirre, J.S., González, A., Özçelik, N., Rodríguez, M.R., García de Fernando, G.D., 2013. Modeling the *Listeria innocua* micropopulation lag phase and its variability. Int. J. Food Microbiol. 164, 60–69. https://doi.org/10.1016/j.ijfoodmicro.2013.03.003

Aguirre, J.S., Pin, C., Rodriguez, M.R., Garcia de Fernando, G.D., 2009. Analysis of the variability in the number of viable bacteria after mild heat treatment of food. Appl. Environ. Microbiol. 75, 6992–6997. https://doi.org/10.1128/AEM.00452-09

Amor, K. Ben, Breeuwer, P., Verbaarschot, P., Rombouts, F.M., Akkermans, A.D.L., De Vos, W.M., Abee, T., 2002. Multiparametric flow cytometry and cell sorting for the assessment of viable, injured, and dead bifidobacterium cells during bile salt stress. Appl. Environ. Microbiol. https://doi.org/10.1128/AEM.68.11.5209-5216.2002

Arku, B., Fanning, S., Jordan, K., 2011a. Flow cytometry to assess biochemical pathways in heat-stressed *Cronobacter* spp. (formerly *Enterobacter sakazakii*). J. Appl. Microbiol. 111, 616–624. https://doi.org/10.1111/j.1365-2672.2011.05075.x

Arku, B., Fanning, S., Jordan, K., 2011b. Heat adaptation and survival of *Cronobacter* spp. (formerly *Enterobacter sakazakii*). Foodborne Pathog. Dis. 8, 975–981. https://doi.org/10.1089/fpd.2010.0819

Baatout, S., De Boever, P., Mergeay, M., 2005. Temperature-induced changes in bacterial physiology as determined by flow cytometry. Ann. Microbiol. 55, 73–80.

Bar-Oz, B., Preminger, A., Peleg, O., Block, C., Arad, I., 2007. *Enterobacter sakazakii* infection in the newborn. Acta Paediatr. 90, 356–358. https://doi.org/10.1111/j.1651-2227.2001.tb00319.x

Baranyi, J., Roberts, T.A., 1994. A dynamic approach to predicting bacterial growth in

food. Int. J. Food Microbiol. 23, 277–294. https://doi.org/10.1016/0168-1605(94)90157-0

Barbesti, S., Citterio, S., Labra, M., Baroni, M.D., Neri, M.G., Sgorbati, S., 2000. Two and three-color fluorescence flow cytometric analysis of immunoidentified viable bacteria. Cytometry 40, 214–218. https://doi.org/10.1002/1097-0320(20000701)40:3<214::AID-CYTO6>3.0.CO;2-M

Ben-Amor, K., Heilig, H., Smidt, H., Vaughan, E.E., Abee, T., de Vos, W.M., 2005. Genetic diversity of viable, injured, and dead fecal bacteria assessed by fluorescence-activated cell sorting and 16S rRNA gene analysis. Appl. Environ. Microbiol. 71, 4679–4689. https://doi.org/10.1128/AEM.71.8.4679-4689.2005

Berney, M., Hammes, F., Bosshard, F., Weilenmann, H.U., Egli, T., 2007. Assessment and interpretation of bacterial viability by using the LIVE/DEAD BacLight kit in combination with flow cytometry. Appl. Environ. Microbiol. 73, 3283–3290. https://doi.org/10.1128/AEM.02750-06

Berney, M., Weilenmann, H.-U., Egli, T., 2006. Flow-cytometric study of vital cellular functions in *Escherichia coli* during solar disinfection (SODIS). Microbiology 152, 1719–29. https://doi.org/10.1099/mic.0.28617-0

Beuchat, L.R., Kim, H., Gurtler, J.B., Lin, L.-C., Ryu, J.-H., Richards, G.M., 2009. Cronobacter sakazakii in foods and factors affecting its survival, growth, and inactivation. Int. J. Food Microbiol. 136, 204–213. https://doi.org/10.1016/j.ijfoodmicro.2009.02.029

Biering, G., Karlsson, S., Clark, N.C., Jonsdottir, K.E., Ludvigsson, P., Steingrimsson, O., 1989. Three cases of neonatal meningitis caused by *Enterobacter sakazakii* in powdered milk. J. Clin. Microbiol. 27, 2054–2056. https://doi.org/10.1046/j.1523-1739.1993.07040815.x

Braux, A.S., Minet, J., Tamanai-Shacoori, Z., Riou, G., Cormier, M., 1997. Direct enumeration of injured *Escherichia coli* cells harvested onto membrane filters. J. Microbiol. Methods 31, 1–8. https://doi.org/10.1016/S0167-7012(97)00073-0

Bunthof, C.J., Abee, T., 2002. Development of a Flow Cytometric Method To Analyze Subpopulations of bacteria in probiotic products and dairy starters. Appl. Environ. Microbiol. 68, 2934–2942. https://doi.org/10.1128/AEM.68.6.2934-2942.2002

Centers for Disease Control and Prevention, 2002. *Enterobacter sakazakii* infections associated with the use of powdered infant formula - Tennessee, 2001. Morb. Mortal. Wkly. Rep. https://doi.org/mm5425a1 [pii]

Chamberlain, J., Moss, S.H., 1987. Lipid peroxidation and other membrane damage produced in *Escherichia coli* K1060 by near-UV radiation and deuterium oxide. Photochem. Photobiol. https://doi.org/10.1111/j.1751-1097.1987.tb07389.x

Chen, C.-Y., Nace, G.W., Irwin, P.L., 2003. A 6×6 drop plate method for simultaneous colony counting and MPN enumeration of *Campylobacter jejuni, Listeria monocytogenes*, and *Escherichia coli*. J. Microbiol. Methods 55, 475–479. https://doi.org/10.1016/S0167-7012(03)00194-5

Choi, M.J., Kim, S.A., Lee, N.Y., Rhee, M.S., 2013. New decontamination method based on caprylic acid in combination with citric acid or vanillin for eliminating *Cronobacter sakazakii* and *Salmonella enterica* serovar Typhimurium in reconstituted infant formula. Int. J. Food Microbiol. 166, 499–507. https://doi.org/10.1016/j.ijfoodmicro.2013.08.016

Davey, H.M., 2002. Flow cytometric techniques for the detection of microorganisms. Methods Cell Sci. https://doi.org/10.1023/A:1024106317540

Edelson-Mammel, S.G., Buchanan, R.L., 2004. Thermal inactivation of *Enterobacter sakazakii* in rehydrated infant formula. J. Food Prot. 67, 60–63.

https://doi.org/10.4315/0362-028X-67.1.60

Farmer, J.J., Asbury, M.A., Hickman, F.W., Brenner, D.J., 1980. *Enterobacter sakazakii*: a new species of "Enterobacteriaceae" isolated from clinical specimens. Int. J. Syst. Bacteriol. 30, 569–584. https://doi.org/10.1099/00207713-30-3-569

Forsythe, S.J., 2005. *Enterobacter sakazakii* and other bacteria in powdered infant milk formula. Matern. Child Nutr. 1, 44–50. https://doi.org/10.1111/j.1740-8709.2004.00008.x

Fredrickson, A.G., Ramkrishna, D., Tsuchiya, H.M., 1967. Statistics and dynamics of procaryotic cell populations. Math. Biosci. 1, 327–374. https://doi.org/10.1016/0025-5564(67)90008-9

Gasol, J.M., Zweifel, U.L., Peters, F., Fuhrman, J.A., Hagström, A., 1999. Significance of size and nucleic acid content heterogeneity as measured by flow cytometry in natural planktonic bacteria. Appl. Environ. Microbiol. 65, 4475–83.

Gregori, G., Citterio, S., Ghiani, A., Labra, M., Sgorbati, S., Brown, S., Denis, M., 2001. Resolution of viable and membrane-compromised bacteria in freshwater and marine waters based on analytical flow cytometry and nucleic acid double staining. Appl. Environ. Microbiol. 67, 4662–4670. https://doi.org/10.1128/AEM.67.10.4662-4670.2001

Gunasekera, T.S., Attfield, P. V., Veal, D.A., 2000. A flow cytometry method for rapid detection and enumeration of total bacteria in milk. Appl. Environ. Microbiol. 66, 1228–1232. https://doi.org/10.1128/AEM.66.3.1228-1232.2000

Herrero, M., Quiros, C., Garcia, L.A., Diaz, M., 2006. Use of flow cytometry to follow the physiological states of microorganisms in cider fermentation processes. Appl. Environ. Microbiol. 72, 6725–6733. https://doi.org/10.1128/AEM.01183-06

Hoefel, D., Grooby, W.L., Monis, P.T., Andrews, S., Saint, C.P., 2003. Enumeration of water-borne bacteria using viability assays and flow cytometry: a comparison to culture-based techniques. J. Microbiol. Methods 55, 585–97. https://doi.org/10.1016/S0167-7012(03)00201-X

International Commission for Microbiological Specification for Foods, I., 2002. Microorganisms in foods 7: Microbiological testing in food safety management, 1st ed. Springer US.

Iversen, C., Lane, M., Forsythe, S.J., 2004. The growth profile, thermotolerance and biofilm formation of *Enterobacter sakazakii* grown in infant formula milk. Lett. Appl. Microbiol. 38, 378–382. https://doi.org/10.1111/j.1472-765X.2004.01507.x

Iversen, C., Mullane, N., McCardell, B., Tall, B.D., Lehner, A., Fanning, S., Stephan, R., Joosten, H., 2008. *Cronobacter* gen. nov., a new genus to accommodate the biogroups of *Enterobacter sakazakii*, and proposal of *Cronobacter sakazakii* gen. nov., comb. nov., *Cronobacter malonaticus* sp. nov., *Cronobacter turicensis* sp. nov., *Cronobacter muytjensii* sp. nov., Cro. Int. J. Syst. Evol. Microbiol. 58, 1442–1447. https://doi.org/10.1099/ijs.0.65577-0

Jordan, K.N., Oxford, L., O'Byrne, C.P., 1999. Survival of low-pH stress by *Escherichia coli* O157:H7: correlation between alterations in the cell envelope and increased acid tolerance. Appl. Environ. Microbiol. 65, 3048–55.

Joshi, S.S., Howell, A.B., D'Souza, D.H., 2014. *Cronobacter sakazakii* reduction by blueberry proanthocyanidins. Food Microbiol. 39, 127–131. https://doi.org/10.1016/j.fm.2013.11.002

Joux, F., Lebaron, P., 2000. Use of fluorescent probes to assess physiological functions of bacteria at single-cell level. Microbes Infect. 2, 1523–1535. https://doi.org/10.1016/S1286-4579(00)01307-1

Kell, D.B., Kaprelyants, A.S., Weichart, D.H., Harwood, C.R., Barer, M.R., 1998.

Viability and activity in readily culturable bacteria: a review and discussion of the practical issues. Antonie Van Leeuwenhoek 73, 169–87. https://doi.org/10.1023/A:1000664013047

Kim, H.T., Choi, H.J., Kim, K.H., 2009. Flow cytometric analysis of *Salmonella enterica* serotype Typhimurium inactivated with supercritical carbon dioxide. J. Microbiol. Methods 78, 155–160. https://doi.org/10.1016/j.mimet.2009.05.010

Koutsoumanis, K., 2008. A study on the variability in the growth limits of individual cells and its effect on the behavior of microbial populations. Int. J. Food Microbiol. 128, 116–121. https://doi.org/10.1016/j.ijfoodmicro.2008.07.013

Lai, K.K., 2001. *Enterobacter sakazakii* infections among neonates, infants, children, and adults. Medicine (Baltimore). 80, 113–22.

Lehner, A., Riedel, K., Eberl, L., Breeuwer, P., Diep, B., Stephan, R., 2005. Biofilm formation, extracellular polysaccharide production, and cell-to-cell signaling in various *Enterobacter sakazakii* strains: aspects promoting environmental persistence. J. Food Prot. 68, 2287–2294. https://doi.org/10.4315/0362-028X-68.11.2287

Lemarchand, K., Parthuisot, N., Catala, P., Lebaron, P., 2001. Comparative assessment of epifluorescence microscopy, flow cytometry and solid-phase cytometry used in the enumeration of specific bacteria in water. Aquat. Microb. Ecol. 25, 301–309. https://doi.org/10.3354/ame025301

Lianou, A., Koutsoumanis, K.P., 2011. Effect of the growth environment on the strain variability of *Salmonella enterica* kinetic behavior. Food Microbiol. 28, 828–837. https://doi.org/10.1016/j.fm.2010.04.006

Manini, E., Danovaro, R., 2006. Synoptic determination of living/dead and active/dormant bacterial fractions in marine sediments. FEMS Microbiol. Ecol. 55,

416–423. https://doi.org/10.1111/j.1574-6941.2005.00042.x

Marty, V., Jasnin, M., Fabiani, E., Vauclare, P., Gabel, F., Trapp, M., Peters, J., Zaccai, G., Franzetti, B., 2013. Neutron scattering: a tool to detect in vivo thermal stress effects at the molecular dynamics level in micro-organisms. J. R. Soc. Interface 10, 20130003–20130003. https://doi.org/10.1098/rsif.2013.0003

Metris, A., George, S.M., Mackey, B.M., Baranyi, J., 2008. Modeling the variability of single-cell lag times for *Listeria innocua* populations after sublethal and lethal heat treatments. Appl. Environ. Microbiol. 74, 6949–6955. https://doi.org/10.1128/AEM.01237-08

Müller, S., Nebe-von-Caron, G., 2010. Functional single-cell analyses: flow cytometry and cell sorting of microbial populations and communities. FEMS Microbiol. Rev. 34, 554–587. https://doi.org/10.1111/j.1574-6976.2010.00214.x

Nebe-von-Caron, G., Stephens, P.., Hewitt, C.., Powell, J.., Badley, R.., 2000. Analysis of bacterial function by multi-colour fluorescence flow cytometry and single cell sorting. J. Microbiol. Methods 42, 97–114. https://doi.org/10.1016/S0167-7012(00)00181-0

Pagotto, F., Nazarowec-White, M., Bidawid, S., Farber, J., 2003. *Enterobacter sakazakii*: infectivity and enterotoxin production in vitro and in vivo. J. Food Prot. 66, 370–376. https://doi.org/10.4315/0362-028X.JFP-11-546

Papadimitriou, K., Pratsinis, H., Nebe-von-Caron, G., Kletsas, D., Tsakalidou, E., 2006. Rapid assessment of the physiological status of *Streptococcus macedonicus* by flow cytometry and fluorescence probes. Int. J. Food Microbiol. 111, 197–205. https://doi.org/10.1016/j.ijfoodmicro.2006.04.042

Parra-Flores, J., Juneja, V., de Fernando, G.G., Aguirre, J., 2016. Variability in cell response of *Cronobacter sakazakii* after mild-heat treatments and its impact on food safety. Front. Microbiol. 7, 1–14. https://doi.org/10.3389/fmicb.2016.00535

Pianetti, A., Falcioni, T., Bruscolini, F., Sabatini, L., Sisti, E., Papa, S., 2005. Determination of the viability of *Aeromonas hydrophila* in different types of water by flow cytometry, and comparison with classical methods. Appl. Environ. Microbiol. 71, 7948–7954. https://doi.org/10.1128/AEM.71.12.7948-7954.2005

Pina-Pérez, M.C., Rodrigo, D., Martínez, A., 2016. Nonthermal inactivation of *Cronobacter sakazakii* in infant formula milk: a review. Crit. Rev. Food Sci. Nutr. 56, 1620–1629. https://doi.org/10.1080/10408398.2013.781991

Raghav, M., Aggarwal, P.K., 2007. Purification and characterization of *Enterobacter sakazakii* enterotoxin. Can J Microbiol 53, 750–755. https://doi.org/10.1139/W07-037

Sachidanandham, R., Yew- Hoong Gin, K., Laa Poh, C., 2005. Monitoring of active but non-culturable bacterial cells by flow cytometry. Biotechnol. Bioeng. 89, 24–31. https://doi.org/10.1002/bit.20304

Saegeman, V.S.M., De Vos, R., Tebaldi, N.D., van der Wolf, J.M., Bergervoet, J.H.W., Verhaegen, J., Lismont, D., Verduyckt, B., Ectors, N.L., 2007. Flow cytometric viability assessment and transmission electron microscopic morphological study of bacteria in glycerol. Microsc. Microanal. 13, 18–29. https://doi.org/10.1017/S1431927607070079

Shapiro, H.M., 2000. Microbial analysis at the single-cell level: tasks and techniques. J. Microbiol. Methods 42, 3–16. https://doi.org/10.1016/S0167-7012(00)00167-6

Simmons, B.P., Gelfand, M.S., Haas, M., Metts, L., Ferguson, J., 1989. *Enterobacter sakazakii* infections in neonates associated with intrinsic contamination of a powdered infant formula. Infect. Control Hosp. Epidemiol. 10, 398–401. https://doi.org/10.1086/646060

Stocks, S.M., 2004. Mechanism and use of the commercially available viability stain, BacLight. Cytometry 61A, 189–195. https://doi.org/10.1002/cyto.a.20069

Subires, A., Yuste, J., Capellas, M., 2014. Flow cytometry immunodetection and membrane integrity assessment of *Escherichia coli* O157:H7 in ready-to-eat pasta salad during refrigerated storage. Int. J. Food Microbiol. 168–169, 47–56. https://doi.org/10.1016/j.ijfoodmicro.2013.10.013

Tehei, M., Franzetti, B., Madern, D., Ginzburg, M., Ginzburg, B.Z., Giudici-Orticoni, M.-T., Bruschi, M., Zaccai, G., 2004. Adaptation to extreme environments: macromolecular dynamics in bacteria compared in vivo by neutron scattering. EMBO Rep. 5, 66–70. https://doi.org/10.1038/sj.embor.7400049

Ueckert, J.E., Nebe von-Caron, G., Bos, A.P., Ter Steeg, P.F., 1997. Flow cytometric analysis of *Lactobacillus plantarum* to monitor lag times, cell divisionand injury. Lett. Appl. Microbiol. 25, 295–299. https://doi.org/10.1046/j.1472-765X.1997.00225.x

van Acker, J., de Smet, F., Muyldermans, G., Bougatef, A., Naessens, A., Lauwers, S., 2001. Outbreak of necrotizing enterocolitis associated with *Enterobacter sakazakii* in powdered milk formula. J. Clin. Microbiol. 39, 293–297. https://doi.org/10.1128/JCM.39.1.293-297.2001

Virta, M., Lineri, S., Kankaanpää, P., Karp, M., Peltonen, K., Nuutila, J., Lilius, E.M., 1998. Determination of complement-mediated killing of bacteria by viability staining and bioluminescence. Appl. Environ. Microbiol. 64, 515–9.

Wagner, S., Snipes, W., 1982. Effects of acridine plus near-ultraviolet light on the outer membrane of *Escherichia coli*. Photochem. Photobiol. 36, 255–258. https://doi.org/10.1111/j.1751-1097.1982.tb04373.x

Wesche, A.M., Gurtler, J.B., Marks, B.P., Ryser, E.T., 2009. Stress, sublethal injury,

resuscitation, and virulence of bacterial foodborne pathogens. J. Food Prot. 72, 1121–1138. https://doi.org/10.4315/0362-028X-72.5.1121

WHO, F. of the U., 2007. Safe preparation, storage and handling of powdered infant formula Guidelines. World Heal. Organ.

Xu, Y.Z., Métris, A., Stasinopoulos, D.M., Forsythe, S.J., Sutherland, J.P., 2015. Effect of heat shock and recovery temperature on variability of single cell lag time of *Cronobacter turicensis*. Food Microbiol. 45, 195–204. https://doi.org/10.1016/j.fm.2014.04.003

Yan, Q.Q., Condell, O., Power, K., Butler, F., Tall, B.D., Fanning, S., 2012. *Cronobacter* species (formerly known as *Enterobacter sakazakii*) in powdered infant formula: a review of our current understanding of the biology of this bacterium. J. Appl. Microbiol. 113, 1–15. https://doi.org/10.1111/j.1365-2672.2012.05281.x

Zhao, X., Zhong, J., Wei, C., Lin, C.-W., Ding, T., 2017. Current perspectives on viable but non-culturable state in foodborne pathogens. Front. Microbiol. 8. https://doi.org/10.3389/fmicb.2017.00580

Figure 1. Flow cytometry plot of SYTO9-PI stained alcohol-treated dead cells of *Cronobacter sakazakii*. The distribution of the observed events was a function of forward and side light scatering. Results from one of the assays carried out in triplicate.

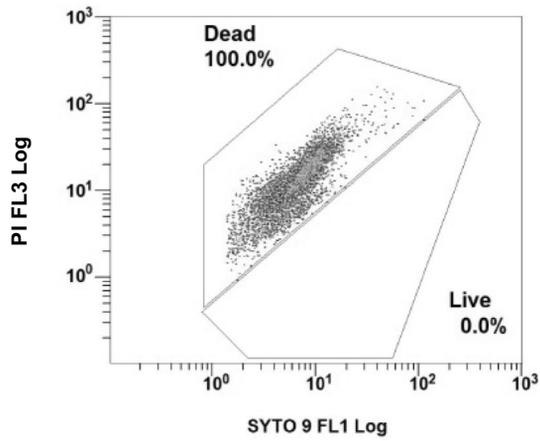

Figure 2. Flow cytometry plots of SYTO9-PI stained *Cronobacter sakazakii* cells from non-treated cell suspensions (nTCS) and treated cell suspensions (TCS; 0% nTCS). Different ratios of nTCS:TCS are shown (0, 25, 50, or 75% nTCS) after heat treatments at 60, 65, or 100 °C. The plots show gates for live, compromised, and dead cells. Results from one of the assays carried out in triplicate.

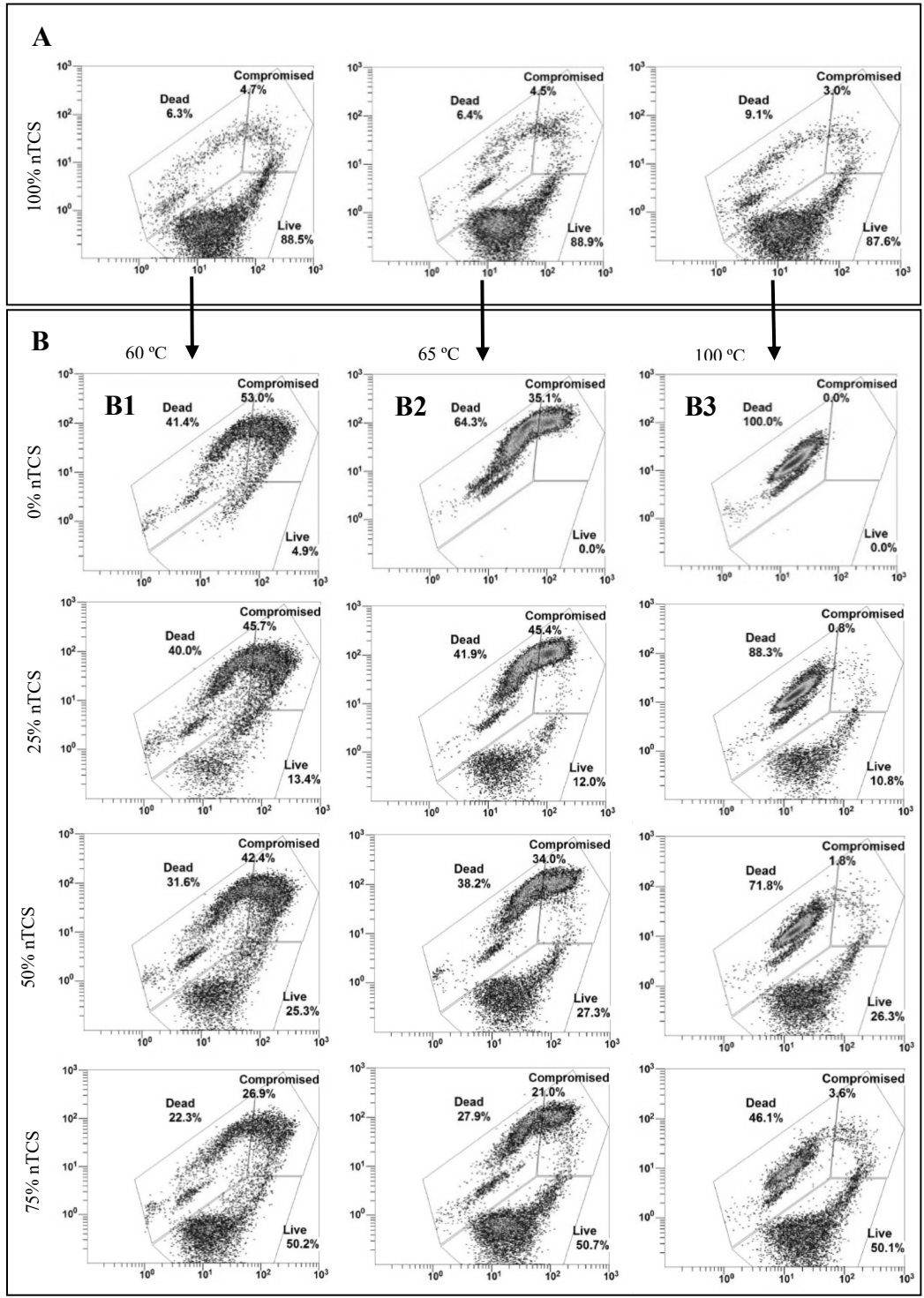

Figure 3. Correlation between agar plate counts (log CFU/ml) of *C. sakazakii* and total bacterial counts detected by flow cytometry (log FC counts/ml).

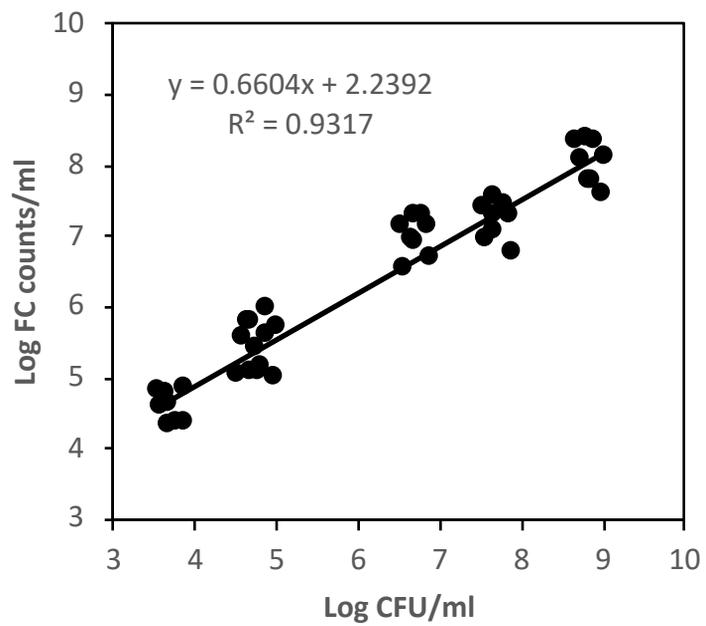